\documentclass[11pt]{article}

\usepackage{paspconf}
\usepackage{epsfig}
\usepackage{natbib}

\def\plotone#1{\centering \leavevmode
\epsfxsize=0.90\columnwidth \epsfbox{#1}}
\def\plottwo#1#2{\centering \leavevmode
\epsfxsize=.45\columnwidth \epsfbox{#1} \hfil
\epsfxsize=.45\columnwidth \epsfbox{#2}}

\begin{document}

\title{Constraints on the Spatial Distribution and Luminosity function of GRBs}

\author{Tomasz Bulik}
\affil{Nicolaus Copernicus Astronomical Center\\
Bartycka 18, 00716 Warsaw, Poland}

\begin{abstract}
The distance scale to gamma ray bursts has been a subject of
scientific debate for almost thirty years. 
Up to the discovery of afterglows only indirect methods 
could be used to constrain the distance scale to this objects.
I review some of these results, and show the current limits on
the spatial distribution and luminosity function of GRBs. 
The results obtained with different methods indicate that
gamma-ray bursts lie at the typical redshifts  $z_{ave}$ between
$1$ and $2$, however there can exist a long tail of the
distribution stretching to higher redshifts. The width
of the GRB luminosity function (the ratio of the intrinsic 
brightness of the brightest to the
dimmest observed burst) estimated from 
Beppo SAX bursts with redshifts is almost $10^3$.
\end{abstract}

\keywords{gamma rays: bursts}

\section{Introduction}

Ever since the discovery of gamma ray bursts \cite{Klebesadel68}
the main questions posed by the astronomers in this field were
what is  the distance scale to these phenomena, and consequently
what is their spatial distribution. For almost three decades the
main tools for probing  the distance scale to gamma-ray bursts
were  studies of  their statistical properties, and model
dependent physical arguments. In fact in the Great Debate in
1995 the main argument for the cosmological distance scale was
the isotropy of GRBs on the sky
\citep{1995PASP..107.1167P,1995PASP..107.1152L}. Apart from the
distribution studies there have been a number of attempts to
measure the distance scale, and the spatial distribution of GRBs
using different methods, based on e.g. the temporal or spectral
studies, searches for gravitational lensing
etc. Since the discovery of afterglows
\citep{1997IAUC.6576....1C} and  measurement of GRB redshifts
\cite{1997IAUC.6676....3M} we have entered a new era in the GRB
research and now we can
probe directly the spatial distribution of GRBs.
The paper is organized as follows: 
in section 2 I analyze the angular distribution of GRBs, 
in section 3 I review the constraints on their spatial
distribution, in section 4  classes of GRBs are discussed, and
the results are summarized in section 5.

\section{Angular distribution}

The angular distribution of gamma ray bursts is very close to
isotropy. In the current BATSE catalogue
\citep{1998hgrb.symp....3M,4B}
we find that the galactic dipole and quadrupole moments are consistent
with the values expected for isotropy when taking into account
the nonuniform exposure.
 The current measured value of the dipole moments in the galactic 
coordinates is $\left<\cos\theta\right> = -0.024\pm 0.014$, with the expected value
equal to $ \left<\cos\theta\right>_{exp} = -.009\pm 0.002$, 
while the quadrupole moment is $\left<\sin^2 b -1/3\right>
=0.0005 \pm 0.0074$, compared to the expected value $ \left<\sin^2 b
-1/3\right>_{exp} = 0.004 \pm 0.002$. 

\begin{figure}

\plottwo{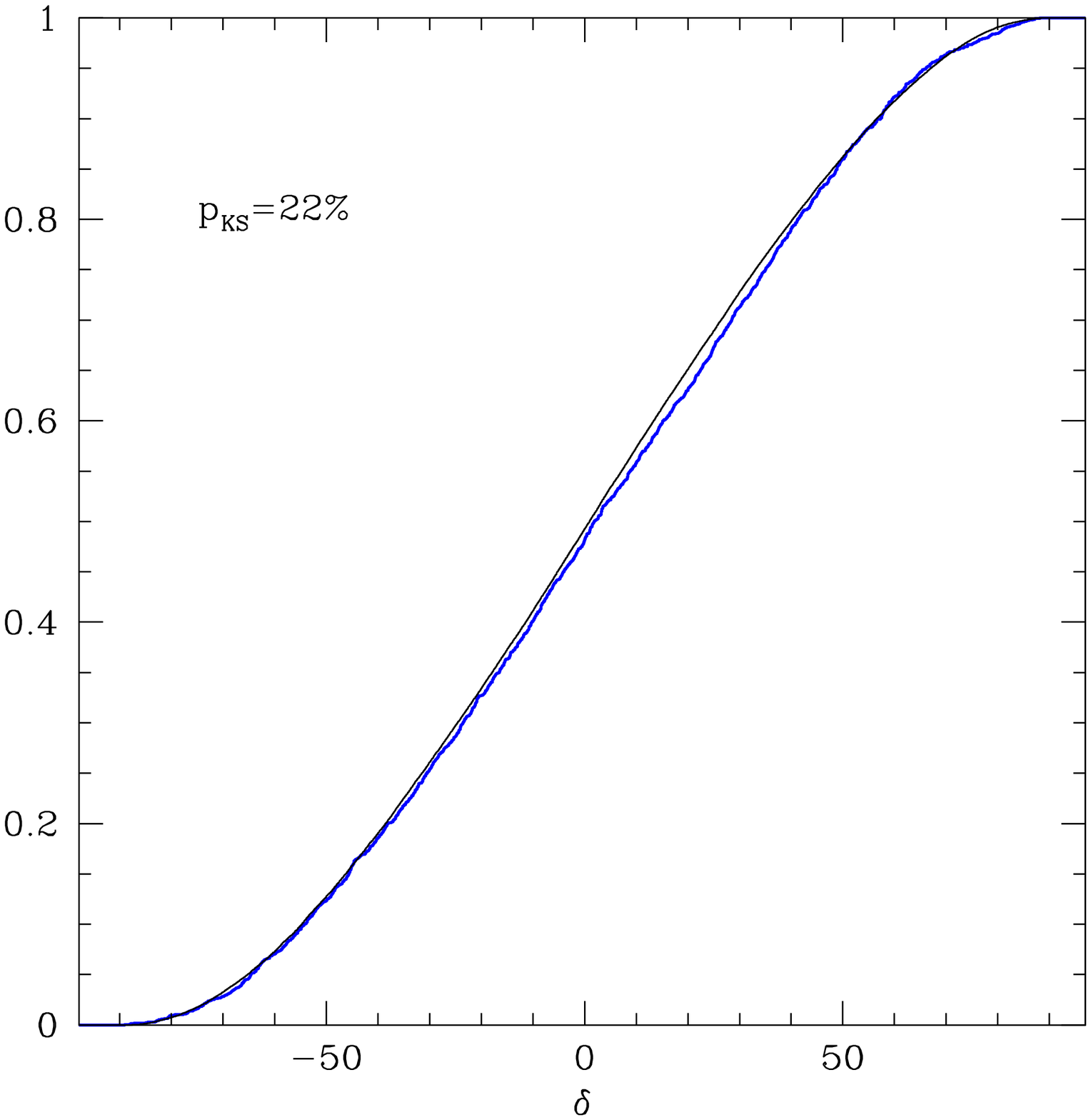}{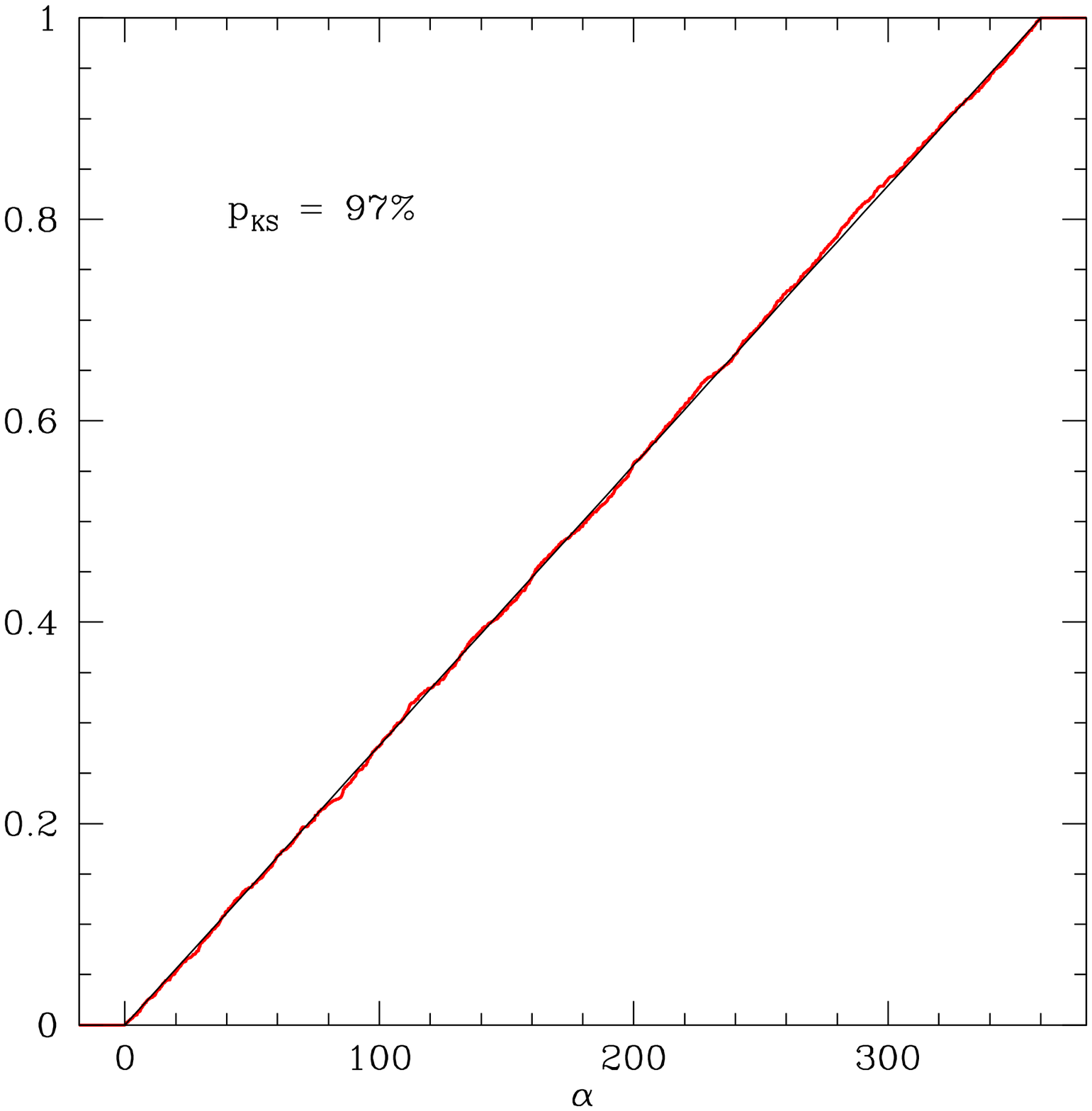}
\caption{Cumulative distributions of the celestial coordinates 
of GRBs in the public archive; 
latitude $b$ on the left, and  longitude $\theta$ on the right.
The thin lines are the distribution expected in the case of
isotropy, and the thick lines are the actual observations.
}
\label{isocel}
\end{figure}

While the integral moments do not describe the entire
distribution, one can perform a more detailed comparison using
the Kolmogorov-Smirnov (KS) tests of the distribution of the galactic coordinates of
the bursts. The KS test comparison shows an excellent agreement
between the expected and observed distributions of right
ascensions $\alpha$, where the probability that the
distributions are identical is 97\%; the probability for the
distribution of declinations is smaller - 22\%. This is probably due to 
uncertainty of the BATSE sky exposure map. This uncertainty is
also seen when comparing the galactic coordinates distribution:
here the KS probabilities are 14\% for galactic latitude
distribution and  32\% for the galactic longitude distribution.

Thus the distribution of GRBs is consistent with isotropy on the
sky. In the first BATSE catalogue \citep{1994ApJS...92..229F}
there was an evidence of variation of the galactic moments with
brightness \citep{1993MNRAS.265L..45Q}. This effect disappeared
in the later catalogues. Therefore in the remaining part of the
paper I assume that the spatial distribution of GRBs can be
factorized into the  independent angular and radial parts,
and that the angular part is isotropic. The most comprehensive
studies of the isotropy  and comparisons with anisotropic models
have been performed by 
\citet{1995ApJS...96..261L,1998ApJ...502..108L}.

\begin{figure}
\plottwo{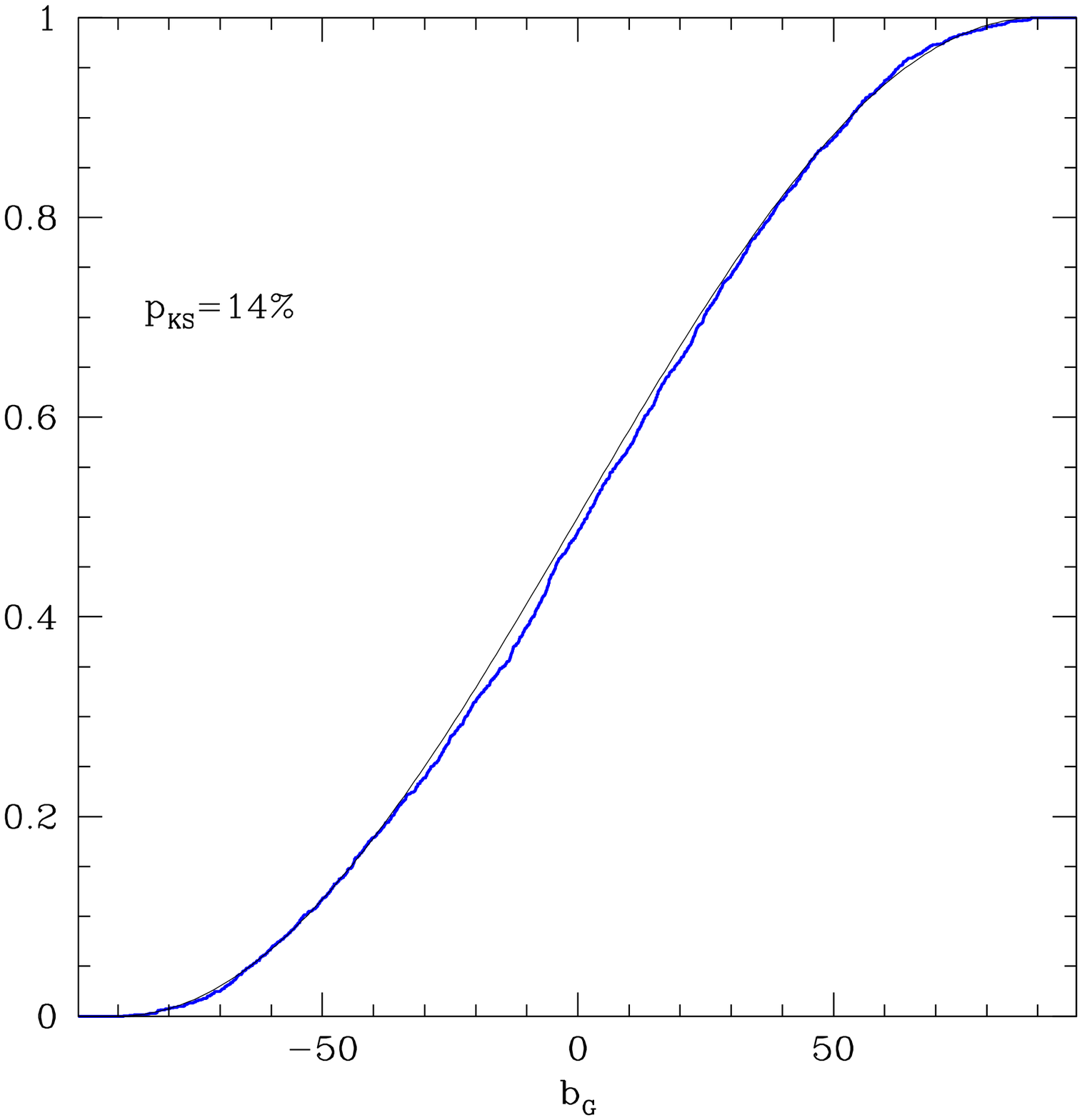}{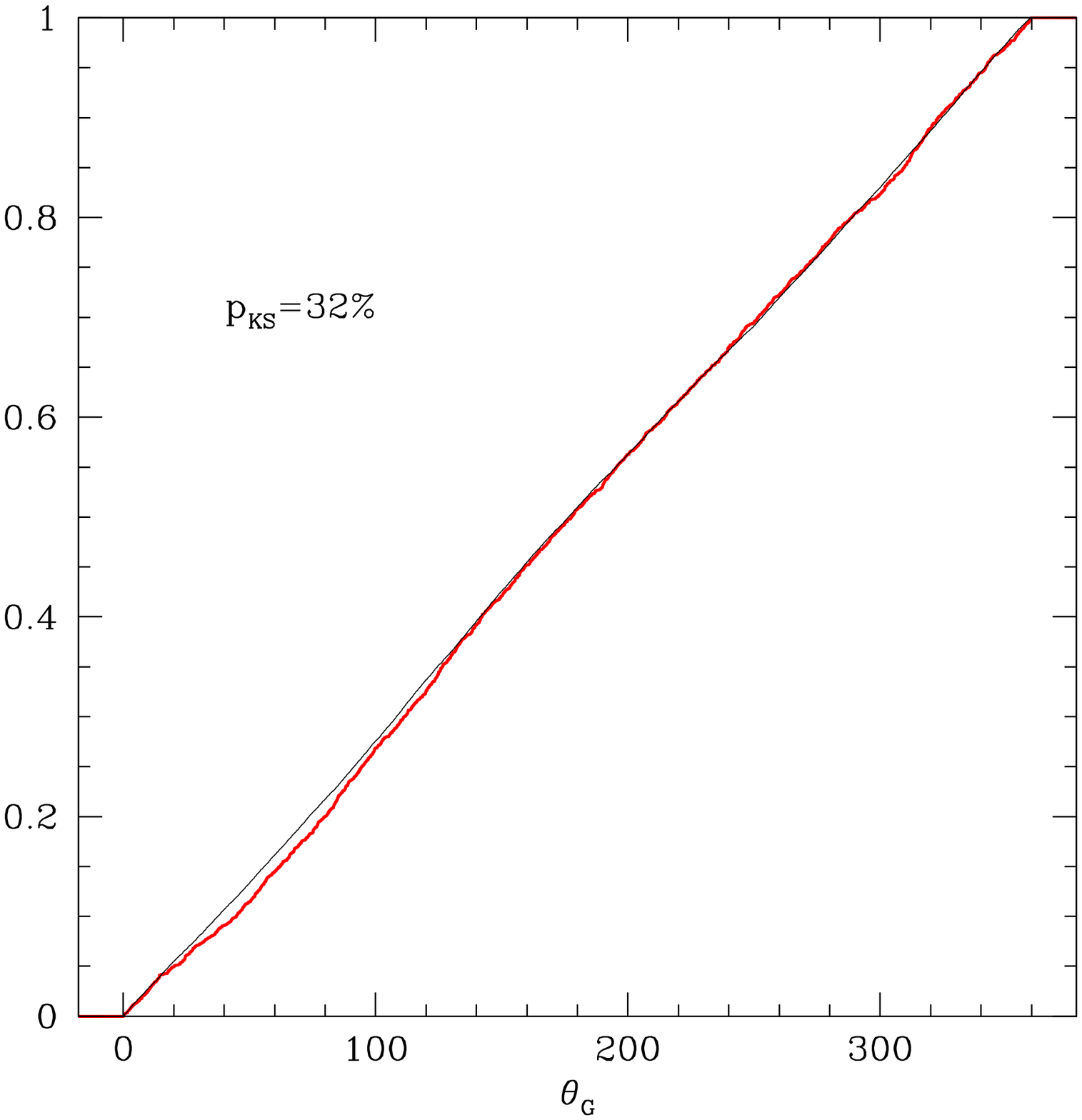}
\caption{Cumulative distributions of the galactic coordinates 
of GRBs in the public archive; 
latitude $b$ on the left, and  longitude $\theta$ on the right.
}
\label{isogal}
\end{figure}

\section{Spatial distribution}

Various statistical methods were used for  analyzing the
gamma-ray burst brightness distributions. Some authors divided
the observed bursts in bins and used a $\chi^2$ statistics to
compare the number of bursts in  these bins with 
models. The shortcoming of this method is that the results
depend on the arbitrary parameter (or parameters), i.e. the
number and  the width of the bins. This makes the calculation of
statistical significance difficult, since in most cases the
authors do not discuss the effects of the bin width. Usually the
bins are chosen in such a way to ensure that the number of
counts in each bin allows using the Gaussian approximation.
Another possibility is to use the Kolmogorov Smirnov test, which
measures the largest deviation between the model and the expected
cumulative distribution.   It seems that the most sensitive, yet
also most computationally demanding method is the maximum
likelihood method, which takes into account all information 
contained in the data.

\subsection{Brightness distribution}

\begin{figure}
\plottwo{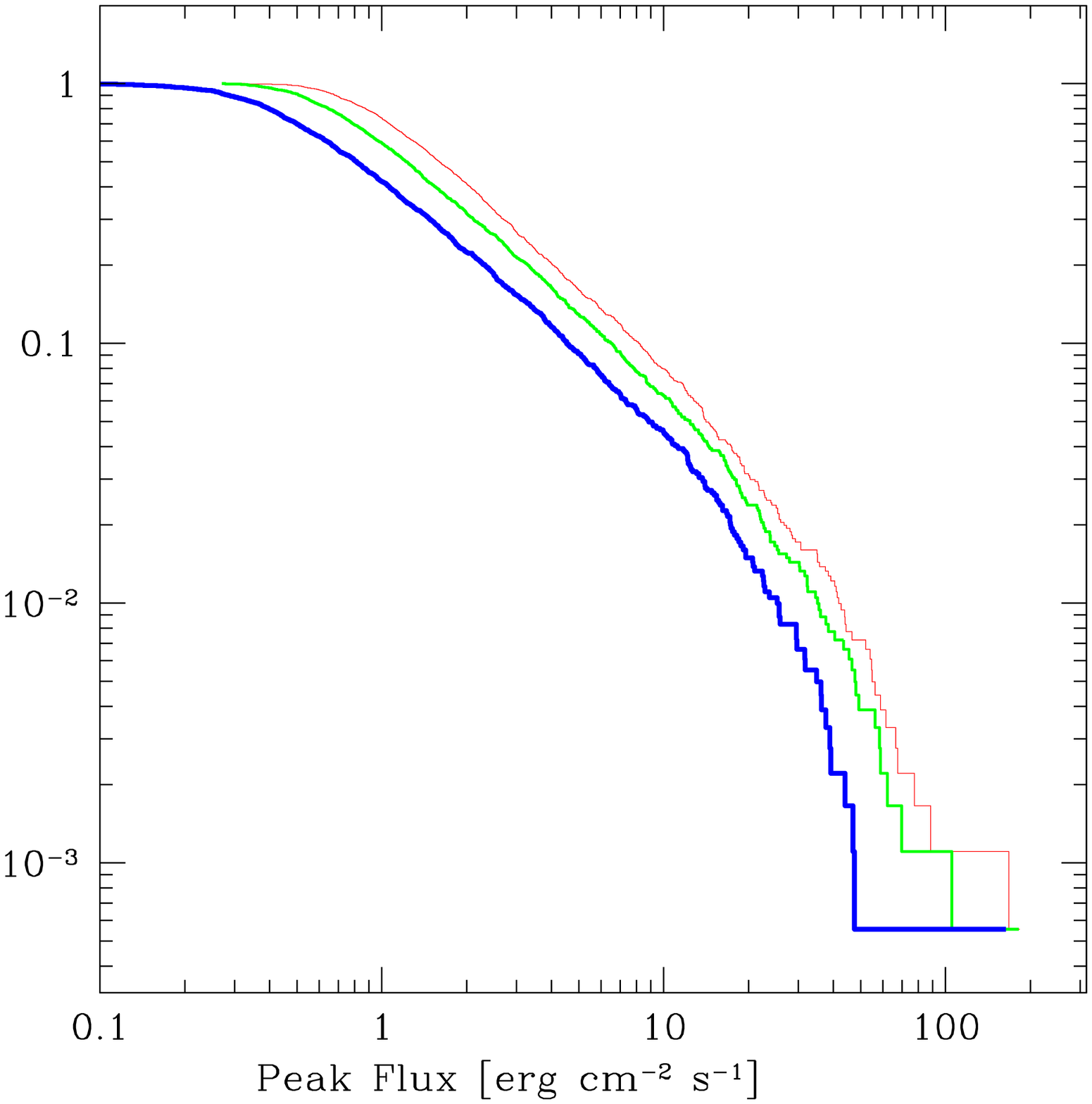}{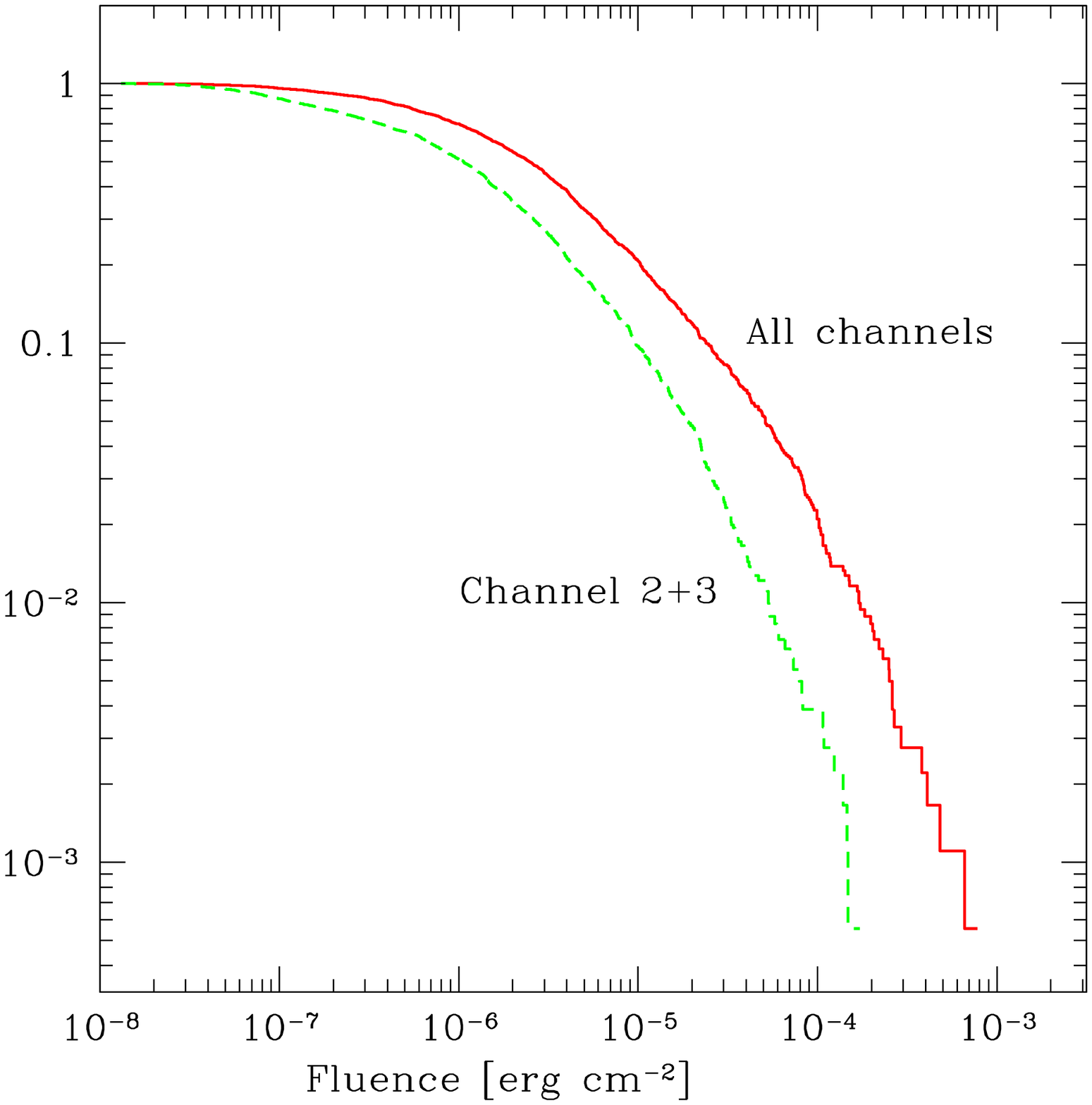}
\caption{Brightness distributions of BATSE GRBs. The left panel
shows the  distributions of peak fluxes on the following
time scales from left to right 64ms (thin line), 256ms (medium
line), and 1024 ms (thick line).
The right panel shows the distribution of peak fluxes: summed
over the BATSE channels (solid line) and for channels 2 and 3
(dashed line).}
\label{bright}
\end{figure}

Studies of GRB brightness distribution require a definition of
brightness of a given burst. Several measures of brightness have
been used: e.g. the peak flux, or the fluence. Each measure of
brightness introduces systematic instrumental effects. The
measured fluence depends on the background level in a particular
detector and on the true time history of a burst. Given that the
fluence of a given burst will not  necessarily follow the $r^{-2}$
law, even for simple, one peak bursts. The
peak flux suffers from the so called Meegan bias. Given a burst
with a couple peaks we will tend to choose the moment of the
maximum flux as the one where the Poisson fluctuation upwards
were the largest. This effect will lead to overestimates of the
flux, and the amount of change will depend on the particular
time history of the burst.  The peak flux distribution will also
depend on the particular timescale on which this peak flux was
measured. The value of each brightness measure depends also on
the spectral response of the detector, and on the spectral
interval considered. The  distributions of peak fluxes, and
fluences of GRBs are shown in Figure~\ref{bright}.

The observed bursting rate $R$ of GRBs can be described by 
\begin{equation}
{dR\over dz\,d\Phi\,dL} ={n(z)\over 1+z} 4\pi r_z^2 {dr_z\over dz}
f(L) \delta(\Phi -\Phi_{det})
\label{basic}
\end{equation}
where $n(z)$ is the density of the sources, $r_z$ is the
luminosity distance, $\Phi$ is the flux, and $f(L)$ is the
luminosity function. In general we measure only the observed flux
$\Phi$, however, recently for some burst also the redshifts $z$
have been measured. Moreover, in a detailed calculation
K-corrections should be added to
equation~\ref{basic}.

The simplest model of cosmological GRBs is with standard candle
luminosity function, and no source evolution. Such a model has
been considered by a number of authors;
\citet{1992ApJ...388L..45M} showed using an $V/V_{max}$ 
 that the brightness
distribution of GRBs is consistent with a simple cosmological
interpretation  with no free parameters. They
 found that the maximum redshift  is constrained
to be $1.0 < z_{max} <2.5$; see also
\citep{1992PhRvL..68.1799D,1992ApJ...389L..45P,1992PhRvL..68.1799D}.
\citet{1993ApJ...411L..55W} found $0.9< z_{max} <2.0$ using the
$C_{max}/C_{min}$ comparison. \citet{1995ApJ...444L..25C} used
the KS-test of the number count distribution in the 2B
catalogue  and found that the maximal redshift must be $1.4<
z_{max} <3.1$. In a detailed study with the use of maximum
likelihood analysis of the 3B data \citet{1998ApJ...502...75L} 
constrained the standard candle luminosity of GRBs to be $L
\approx 0.74\times 10^{51}$erg\,s$^{-1}$
\nocite{1995ApJS...96..261L}. 

The effects of the GRB rate density evolution are usually
included by assuming that  $n(z) \propto (1+z)^D$. In general
the  results show that the higher the exponent $D$, the more
luminous GRBs are. The luminosity of a standard candle bursts
grows very rapidly with the exponent $D$, however the details
vary from  one paper to another and depend on the particular
data set used (BATSE or BATSE and PVO) 
\citep{1995ApJ...444L..25C,1996ApJ...466...29M,1996ApJ...470...56H,1998ApJ...502...75L}

\subsubsection{Luminosity function}

When considering the luminosity function of GRBs one has to 
distinguish between the observed and intrinsic luminosity
functions. The intrinsic luminosity function is the distribution
of luminosities of all bursts regardless of whether we see them
or not, while the observed luminosity function is the
distribution of luminosities of  the observed bursts only. It has
been shown that the two distributions are different, and may
even have different slopes. The luminosity function is usually
represented by a power on a fixed interval, however other
functional forms like e.g. the lognormal distribution are also
used.

There were several papers discussing  the effects of the
luminosity function on the shape of the brightness
distribution.  
\citet{1994ApJ...426L...5H} used the integral moment analysis and
concluded that the width of the observed luminosity function
must be smaller than a factor of $6.5$.
\citet{1995ApJ...444L..25C} used a two $\delta$ functions
approximation of the luminosity function and constrained the
width of such luminosity function to less than a factor of $14$
in the case of BATSE data and less than a factor of $2$ for the
combined BATSE and PVO data sets.  \citet{1995ApJ...453..583W}
parameterized the  luminosity function as a lognormal
distribution, used KS-test to compare with the observed
$C_{max}$ distribution, and found that the width of the
luminosity function must be less than about a factor  of $40$.
Some studies of the effects of the luminosity function were done
assuming a halo-core like spatial distribution: 
\citet{1995ApJ...439..303U} modeled the BATSE brightness
distribution using the maximum likelihood method and found
essentially no limits on the intrinsic luminosity function,
however \citet{1995ApJ...440L...9U} used the combined BATSE+PVO)
dataset and found the the observed luminosity function width is
less than 10. \citet{1998ApJ...505..666B} also considered a
halo-like
distribution and using the KS-test found that BATSE data is
consistent with the luminosity function wider than a factor of
100. Analyses of the combined BATSE+PVO dataset with $\chi^2$ yielded
constraints that the luminosity function width must be less than a
factor of 1000
\citep{1996ApJ...470...56H,1996ApJ...462..125H}. 
\citet{1998ApJ...502...75L} analyze the gamma ray burst
luminosity function using Bayesian statistics and maximum
likelihood method. They find that the current data set (BATSE 3B
catalogue) does not constrain the intrinsic luminosity function.
The main reason is that the differences between the models should
show only for much fainter fluxes than the BATSE threshold
(see Fig. 25 in their paper) . They
also find the width of the {\em observed} luminosity function is 
around $10^4$. 

Thus the studies of the GRB luminosity function by analysis of
the brightness distribution has taken an unusual direction:
initially with small amount of data it seemed that the width of the
luminosity function is small, yet with the increasing amount of
data and improved statistical methods it appeared that its width
can actually be large. There are no strong constraints on the
width of the intrinsic luminosity function of GRBs from the
brightness distribution.

\subsection{Temporal studies}

\begin{figure}
\plottwo{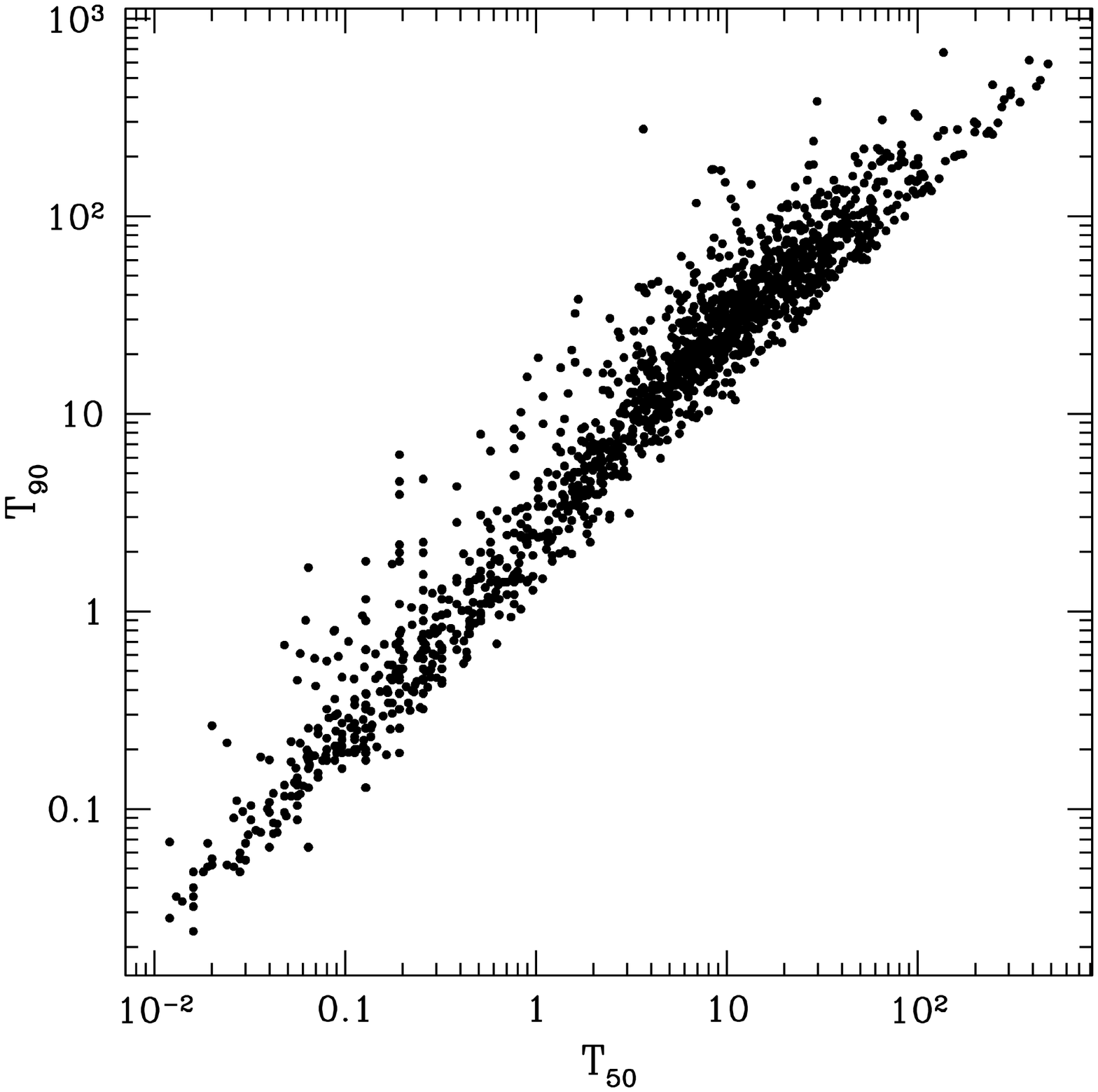}{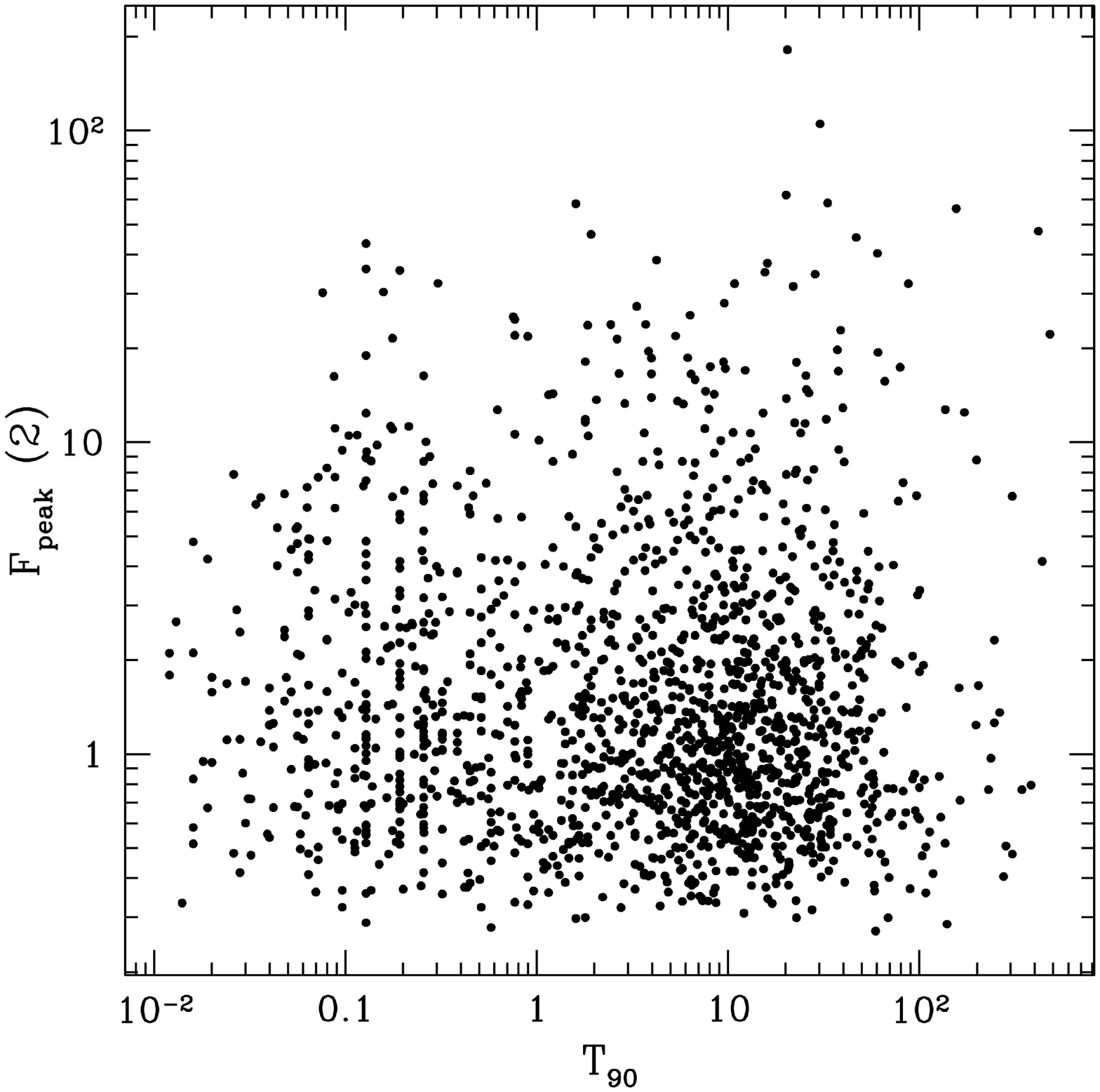}
\caption{The left panel shows the distribution of BATSE
measured burst durations $T_{50}$ and $T_{90}$. While a clear
correlation between these two measures exists there is also a
significant spread. The right panel shows the BATSE bursts in the
plane spanned by $T_{90}$ and the peak flux on the 256ms
timescale. No clear correlation exists, which shows how difficult
it is to find the burst distance scale from brightness duration
correlation.}
\label{temporal}
\end{figure}

It is natural to attempt to use the temporal characteristic of 
GRBs to find constraints on the distance scale 
within the framework of the cosmological model. Assuming that
weaker bursts are typically further away they should also be
longer, due to cosmological redshift. It should be noted that it
is difficult to obtain an absolute scale without a physical model
or knowledge of a clock in bursts. One can only measure the
relative distance (redshift) between groups  of weak and bright
bursts. Thus the absolute distance scale can be obtained then
assuming a distance to the weakest bursts. The method 
is based on finding a correlation between brightness and
duration, or other measure of the intrinsic timescale in
gamma-ray  burst.

Studies of temporal characteristics of GRBs  require  defining a
fundamental timescale for GRBs. This is not easy, since GRB time
histories are very diverse. BATSE catalogue provides two
measures of duration $T_{50}$, and $T_{90}$, the time in which 
$50$ and $90$ percent of the photons arrived.  I present the
correlation between these two measures of duration in left panel
of Figure~\ref{temporal}. The spread in these correlation is
significant and this shows that there are possible systematic
effects that may affect the brightness - duration studies. 
The right panel of Figure~\ref{temporal} shows the distribution
of BATSE GRBs in the duration ($T_{90}$) and brightness (peak
flux on the $256\,$ms timescale) plane. No correlation is
apparent in this graph. Thus
several authors decided  to analyze the BATSE data themselves
and measure durations, or time scales defined differently.

We present a sample of the results obtained with the use of
different methods, different measurements of burst durations in
Table~\ref{fludur}. Typically regardless of the method  a
stretch factor of approximately 2 was found, meaning that the
typical clock for weak bursts runs slower than that for the 
bright ones. There are some exceptions:
\citet{1996ApJ...459..570M} and \citet{1997ApJ...474...37L}
found no difference between the bright and the dim bursts. The 
stretch factor of $\approx 2$ was interpreted as the redshift of the dim
burst of approximately $z_{dim} =2$. However,
\citet{1995ApJ...453...25F} noticed that in each given burst the
width of the peaks is larger when measuring low energy channels
than in the case of high energy ones. This leads to the
so-called W-correction; the effects of redshift is not only
slowing down the clocks but also shifting to lower energies,
where the peaks are wider.  The inclusion of this correction
lead to a new limit  on the redshift of the dim bursts
$z_{dim}\approx 6$ for the stretch factor of $2$. It has to be
noted that  it is always difficult to separate the effects of
source evolution  from the true cosmological effect in  these
studies. In fact \citet{1997ApJ...489L..41S} interpreted the
variation in the slope of decay of GRB profiles as due to the
evolution of sources.

\begin{table}[t]
\caption{GRB time stretching results.} 
\vspace{4mm}

\label{fludur}
\begin{center}
\begin{tabular}{llr}
\hline
Method       & $S$        &  reference  \\
\hline\hline
Peak align   & 2.25       & \citet{1994ApJ...424..540N} \\
Fluence/Peak & 2.0        & \citet{1994ApJ...424..540N}\\
Wavelets     & 2.25       & \citet{1994ApJ...424..540N}\\
Pulse width  & 1.8        & \citet{1995PhDT........17D} \\
Counts in peak & 2.2      & \citet{1995PhDT........17D} \\
Total counts & 2.0        & \citet{1995PhDT........17D}  \\
Duration     & 2.2        & \citet{1995ApJ...439..542N}\\
Autocorrelation & 2.0     & \citet{1995ApJ...448L.101F}\\
Ave. time hist. & 1.0     & \citet{1996ApJ...459..570M}\\
Stretch exponential & 2.0 & \citet{1996ApJ...464L.111S}\\
Fluence vs $T_{50}$ &1.0   & \citet{1997ApJ...474...37L}
\end{tabular}
\end{center}
\end{table}

\subsection{Host galaxy limits}

The host galaxy problem, for quite a while called the "no host
problem" was first presented by \citet{1992..Schaeffer}. He analyzed
the contents of error boxes of brightest bursts with the accurate IPN
localization, in search of
galaxies and found that these boxes contain no galaxies to the
limit of twentieth magnitude, while expecting to find nearby
galaxies of magnitude $\approx 16$ in them.  Of course, in this
calculation  it was 
assumed that GRBs are not totally exotic phenomena and therefore
are taking place in galaxies. This finding led to the conclusion
that the standard cosmological model (with typical GRB
luminosity $6\times 10^{50}$ergs) has some difficulty. An
analysis of the infrared galaxy catalog by \citet{1997ApJ...491...93L},
was claimed to show  consistency of the contents of GRB errors
boxes with the cosmological model, however \cite{1999ApJ...514..862B} 
showed using Bayesian analysis that the infrared data set
provided no really useful limits. 

Recently this problem has been revisited by \citet{Schaeffer98}
and \citet{1999ApJ...514..862B}. \citet{Schaeffer98} analyzed the
contents of 26 small GRB error boxes and concluded that 
the lower limit on GRB luminosity is $6\times
10^{58}\,$phot\,cm$^{_2}\,$s$^{-1}$, assuming no evolution of
sources. \citet{1999ApJ...514..862B} applied the Bayesian methods to the
dataset of \citet{Schaeffer98} and eight BeppoSAX bursts with
good localization and host galaxies. This analysis rules out 
burst energies below $2\times 10^{52}\,$ergs, and favors,
although not strongly, the value $\approx 10^{53}\,$ergs.

\subsection{Gravitational lensing searches}

Given that GRBs are located at the cosmological distances, some
of them should, like all other distant  sources, be
gravitationally lensed \citep{1986ApJ...308L..43P}. Since GRBs are
transients, a lensed burst would consist of two (or more) bursts
with identical spectra and time profiles, however delayed in
time by typically a month. \citet{1992ApJ...389L..41M} first
estimated of the lensing probability to lie between 0.05\% and
0.4\%. The lensing probability decreases with the time delay on
a timescale of a month, and it was found to peak at $\Delta t
\approx 50 s(M_*/10^6 M_\odot)$, where $M_*$ is the mass of the
lensing object. Thus acquiring a large enough database of bursts
should allow to find lensed gamma-ray bursts. However one has to
bear in mind that there are some significant selection effects
against detecting lensed events. First, BATSE does not monitor
all sky simultaneously, the average exposure being about 40\%
of the sky time. Moreover, after detecting a burst BATSE turn
into inactive mode for a 90 minutes. If a burst happens at
this time it is only recorded if it is stronger than the
preceding burst however, than the information on the preceding
burst is lost. Thus detecting lensed bursts
with time delays up to 90 minutes is hardly possible.
Searches for lensed GRB in the BATSE database were
conducted \cite{1994ApJ...432..478N,1998hgrb.symp..166M}
yielding null results.

The gravitational lensing problem has recently been revisited by 
\citet{1999ApJ...510...54H}, who found that actually including
multiple lensing is important. They have calculated lensing
probabilities as a function of the average redshift $z_{ave}$ of the
GRB distribution for different cosmologies. Basing on no
detection of GRB lensing they derive an upper limit on the
average redshift $z_{ave} < 2.2, 2.8, 4.3, 5.3$ for
$(\Omega, \Lambda)$ pairs $(0.3, 0.7)$, $(0.5, 0.5)$,
$(0.5, 0.0)$, and $(1.0,0.0)$ respectively. It should be noted
that the gravitational lensing probes the spatial 
distribution  of GRBs  directly and is not very sensitive
to the luminosity function.

\begin{figure}
\plottwo{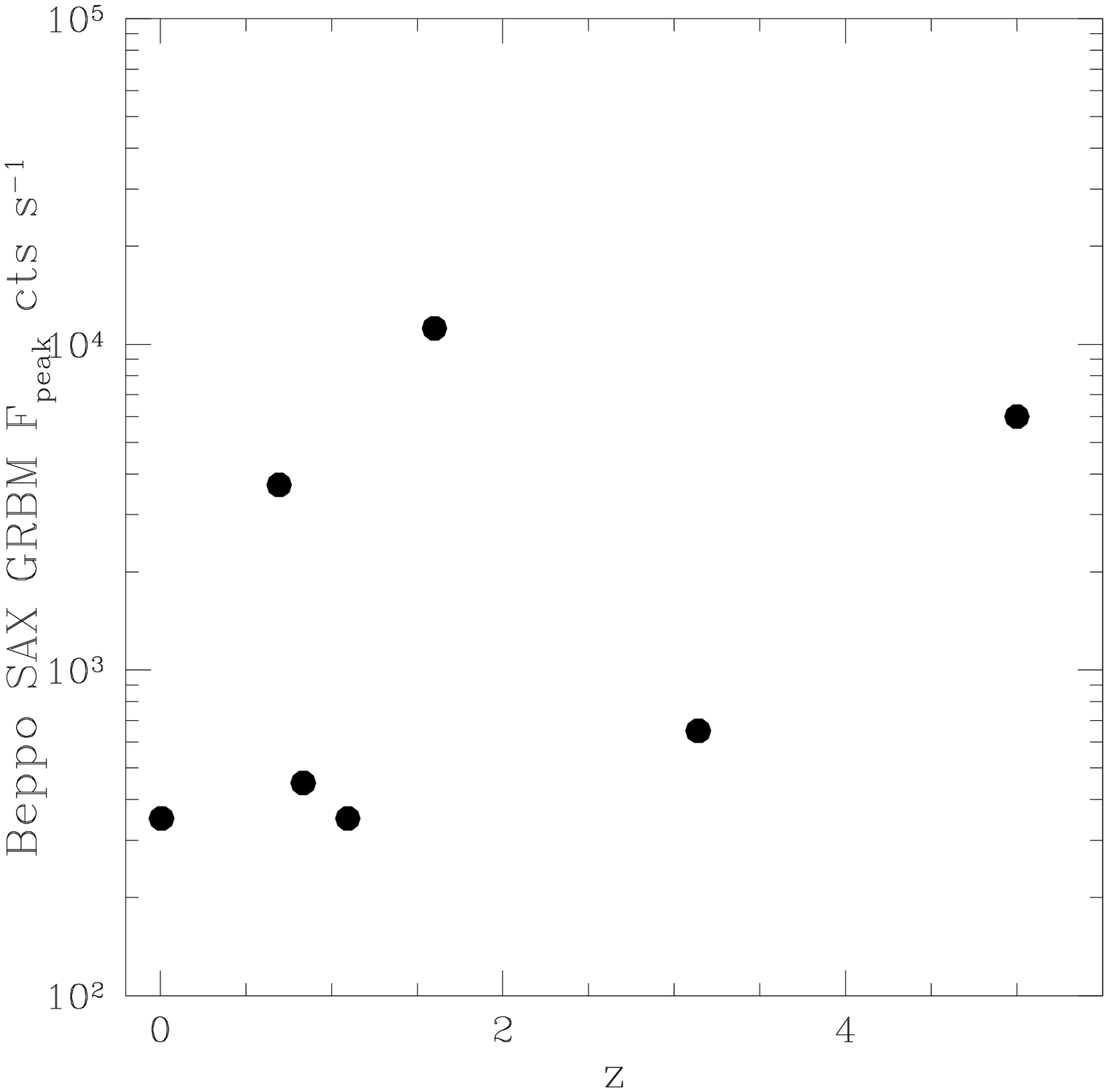}{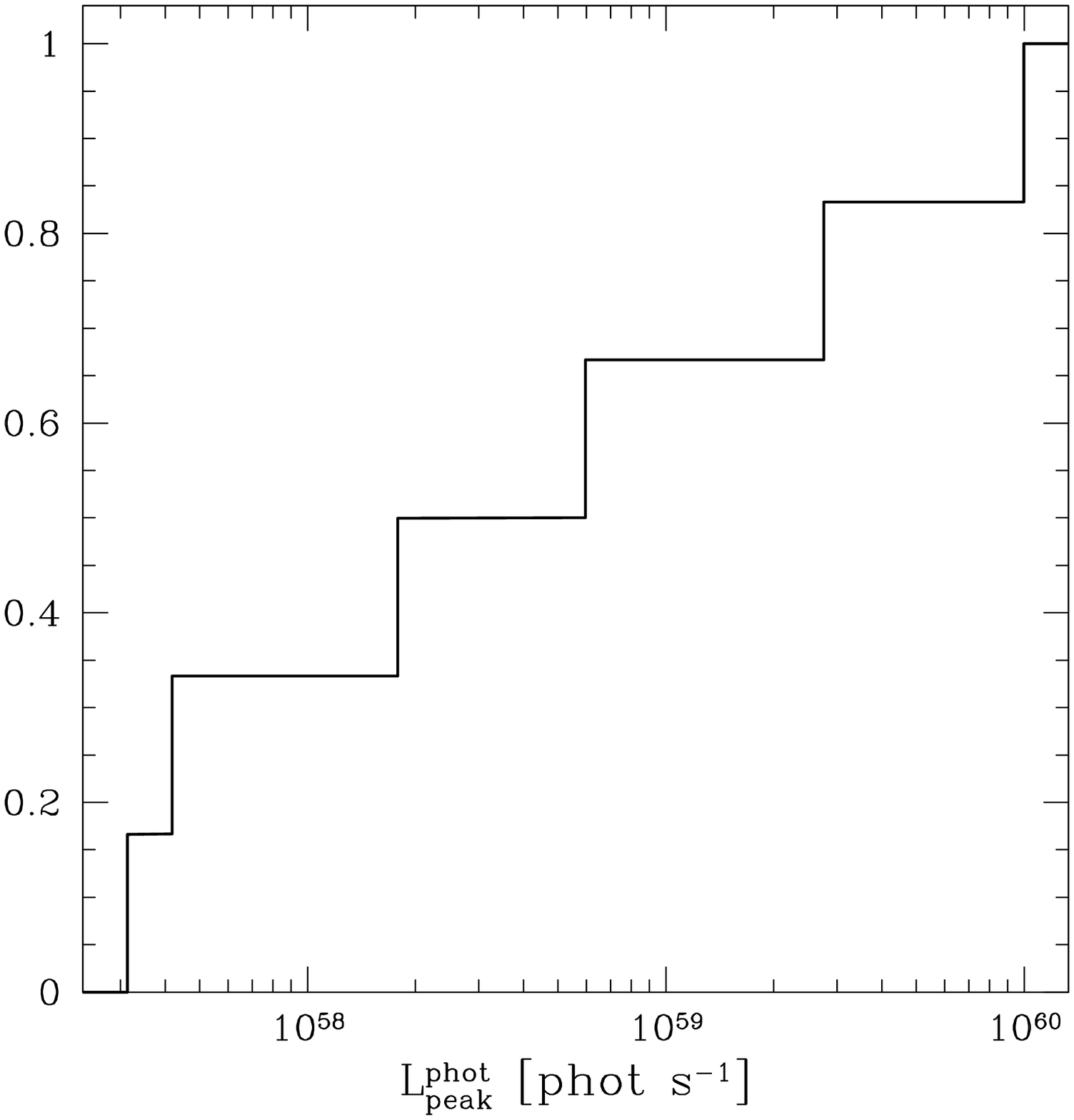}
\caption{The left panel shows seven Beppo SAX burst with measured
redshifts
in the $z$ vs peak flux plane. If GRBs were standard candles
the points would lie on a straight line.
The right panel shows the cumulative distribution of 
peak photon luminosities for SAX bursts with redshifts.
I have omitted  GRB980425. The observed luminosity function
is almost a factor of $10^3$ wide. }
\label{red}
\end{figure}

\subsection{Clustering}

While the overall distribution of GRBs shows an amazing degree
of isotropy there is also a question of isotropy or clustering
on the small angular scales. There has been evidence for such
anisotropy - small scale clustering of burst - in the first
BATSE catalogue found by \citet{1993MNRAS.265L..59Q}, who
interpreted it as evidence for burst repeating. However, with
the change of the BATSE  localization algorithm the small scale
clustering detection  was not confirmed, and has not been seen
in the larger data sets.

Within the framework of the cosmological model small scale
clustering is expected, if bursts trace luminous matter. 
\citet{1993ApJ...415L...1L} suggested that given a large enough
sample of bursts the large scale structure in the Universe
should be detectable: 1000 bursts should suffice to probe the
supercluster scales, and about 3000 bursts should probe  the
scales above $25\,h^{-1}\,$Mpc. These results were used with the
BATSE 3B catalogue to find that the median distance to the weak
bursts must be larger than $630 h^{-1}\,$Mpc, and the closest
burst must be further away than $40 h^{-1}\,$Mpc
\citep{1996ApJ...461L..69Q}. While these limits are not
very constraining they grow approximately linearly with the
number of bursts detected provided that no clustering is
found. Alternatively, the number  of GRBs in the current BATSE
database should soon allow to probe the large scale structure.

\subsection{Direct measurements}

The studies of GRB spatial distribution have entered a new era
in the beginning of 1997 with the BeppoSAX discovery of X-ray
afterglows \citep{1997IAUC.6576....1C}, This led to rapid optical
followups and the discovery of optical
afterglows \citep{1997IAUC.6584....1G},  and consequently to
identification of host galaxies \citep{1997IAUC.6588....1G}. 
The accurate optical spectroscopy of either the afterglows
themself of the host galaxies allows to find GRB redshifts,
with the first redshift of 0.835 measured for GRB970508 
\citep{1997Natur.387..878M}.
At the time of writing there are nine GRB redshifts measured,
including two which are rather uncertain. A list of the
GRBs with measured redshifts is presented in Table~\ref{zees}.
This table was prepared with the aid of the information posted 
on Jochen Greiners web page \citep{Greiner}.

\begin{table}[t]
\caption{GRBs with measured redshifts.}
\vspace{4mm}

\label{zees}
\begin{center}
\begin{tabular}{llrlr}
GRB      & z     & $F_{peak}$\, & $F_{WFC}$  & Reference \\
         &       &cts\,s$^{-1}$ & [Crab]     &            \\
\hline \hline 
970228   & 0.695 &  3700 &0.23 &\citet{1997IAUC.6572....1C}  \\
970508   & 0.835 &  450  & 1.0 & \citet{1997IAUC.6649....1C}\\
970828   & 0.3$^b$&      &0.735$^a$&\citet{1997IAUC.6726....1R}\\
971214   & 3.42  & 650   &  1.0 &\citet{1997IAUC.6787....1H}\\
980329   & 5\,$^c$ &  6000 &  6.0 & \citet{1998IAUC.6853....1F} \\
980425   & 0.008$^d$ &   350 &  3.0 & \citet{1998IAUC.6884....1S} \\
980613   & 1.096 &  350  & 0.6&\citet{1998IAUC.6938....1S}\\
980703   & 0.96  &       & 1.7$^a$
&\citet{1998IAUC.6966....1L}\\
990123   & 1.60  & 11200 & 3.4
&\citet{1999IAUC.7095....1F}\\ \hline
\multicolumn{5}{p{12cm}}{$^a$ RXTE-ASM fluxes between $2$ and
$12\,$keV}\\
\multicolumn{5}{p{12cm}}{$^b$Redshift of $0.3$ is based on a weak
detection of iron line \citep{Yoshida99}; interpreting the  spectral feature
as the iron edge the redshift becomes $\approx 0.9$.}\\
\multicolumn{5}{p{12cm}}{$^c$Redshift determined from broad band
photometry \citep{1999ApJ...512L...1F}. 
A faint  galaxy at the same location has been found, however
it lies  at a lower redshift.}\\
\multicolumn{5}{p{12cm}}{$^d$Redshift determined on the
assumption that SN1998bw is the counterpart of the GRB
\citep{1998Natur.395..670G}.}\\
\end{tabular}\end{center}
\end{table}

We present the "Hubble" diagram for GRBs in the left panel of 
Figure~\ref{red}. There is no clear correlation between the
redshift and peak flux, I even had to use the logarithmic scale
to show all the fluxes. This indicates that the GRB luminosity
function is broad. In the right panel of Figure~\ref{red} I
present the cumulative distribution of peak photon luminosities
for Beppo SAX bursts with redshift. In this Figure
I have omitted two ASM/RXTE
bursts from Table~\ref{zees} and also GRB980425 which, if the
identification with SN1998bw is correct would be five orders of
magnitude fainter. The luminosities have been calculated assuming
that $\Omega_M=0.3$ and $\Omega_\Lambda=0.7$, and
$H_0=65\,$km\,s$^{-1}\,$Mpc$^{-1}$. The observed
luminosity function is a factor of almost $10^3$ wide, and this
result has been obtained already  with a small sample of a few
burst. This is an indication that the intrinsic luminosity of
function of GRBs could be even broader than that!

\begin{figure}[t]
\plotone{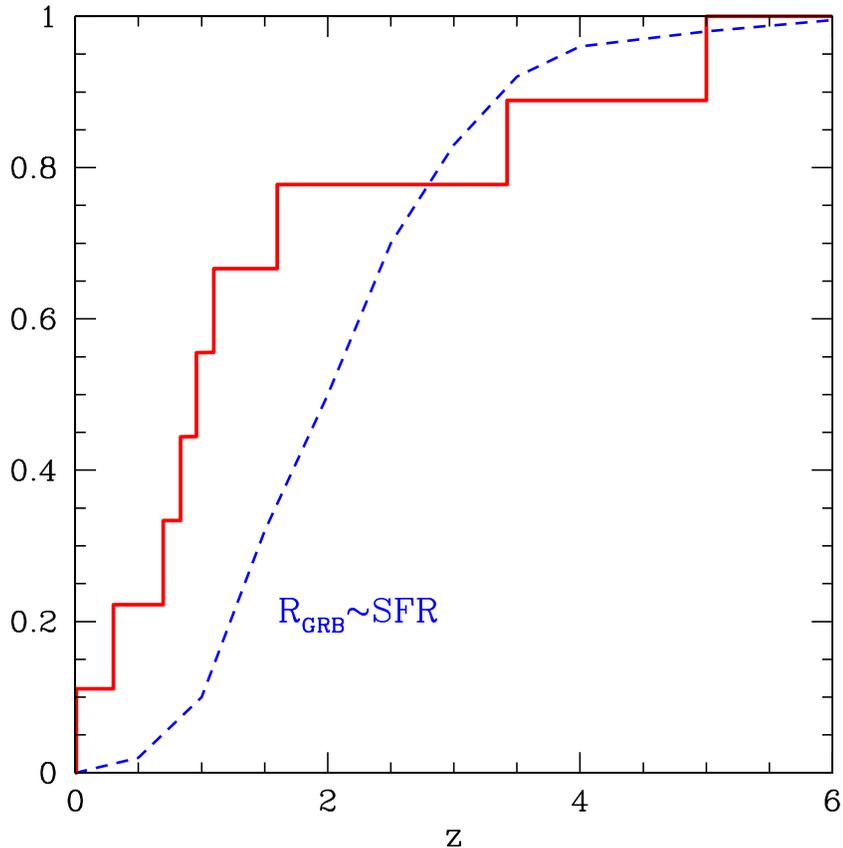}
\caption{Cumulative
 spatial distribution of GRBs from measured redshifts.
 All entries from Table~\ref{zees} are shown.We present  a model 
 where the GRB rate is proportional to star formation 
 rate (dashed line) \citep{1998MNRAS.294L..13W}.} 
\label{zdist}
\end{figure}

With the measurements of redshifts we can now directly 
probe the spatial distribution of GRBs. The cumulative
distribution of GRB redshifts measured to date is presented
in Figure~\ref{zdist}. I have included all the entries from
Table~\ref{zees} in this Figure, despite the fact that, as
discussed above, 
some of the entries
could be uncertain. Also since we know that the luminosity
function is very broad we may not be detecting all the bursts
from high redshifts. Therefore studies of GRBs with more
sensitive instruments are very important. 

Models for GRBs associate them with late stages of 
stellar evolution, either compact object mergers, or collapsars.
Thus one expects a relation between the star formation rate and
the GRB bursting rate. The star formation rate has recently been
measured in the Hubble Deep Field \citep{1996MNRAS.283.1388M},
which shows a dramatic change in the star formation rate between 
$z=1$ and present.  \citet{1997ApJ...486L..71T} calculated
the expected GRB rate based on the known star formation history
and the synthesis of stellar (and binary) evolution. 
\citet{1998MNRAS.294L..13W} fitted the combined BATSE and PVO
brightness distribution with a star formation model, and  the
dimmest burst could come from the redshifts as high as $z\approx
6$. Based on their results I present a model with the GRB rate
proportional to star formation rate in Figure~\ref{zdist}.
While the two curves do not match very well, they look similar
despite the fact that no selection effects due to detector
sensitivity, or luminosity function were taken into account. 
This could be an  indication that GRBs are connected
with star formation.

\section{Classes of GRBs}

\begin{figure}
\plottwo{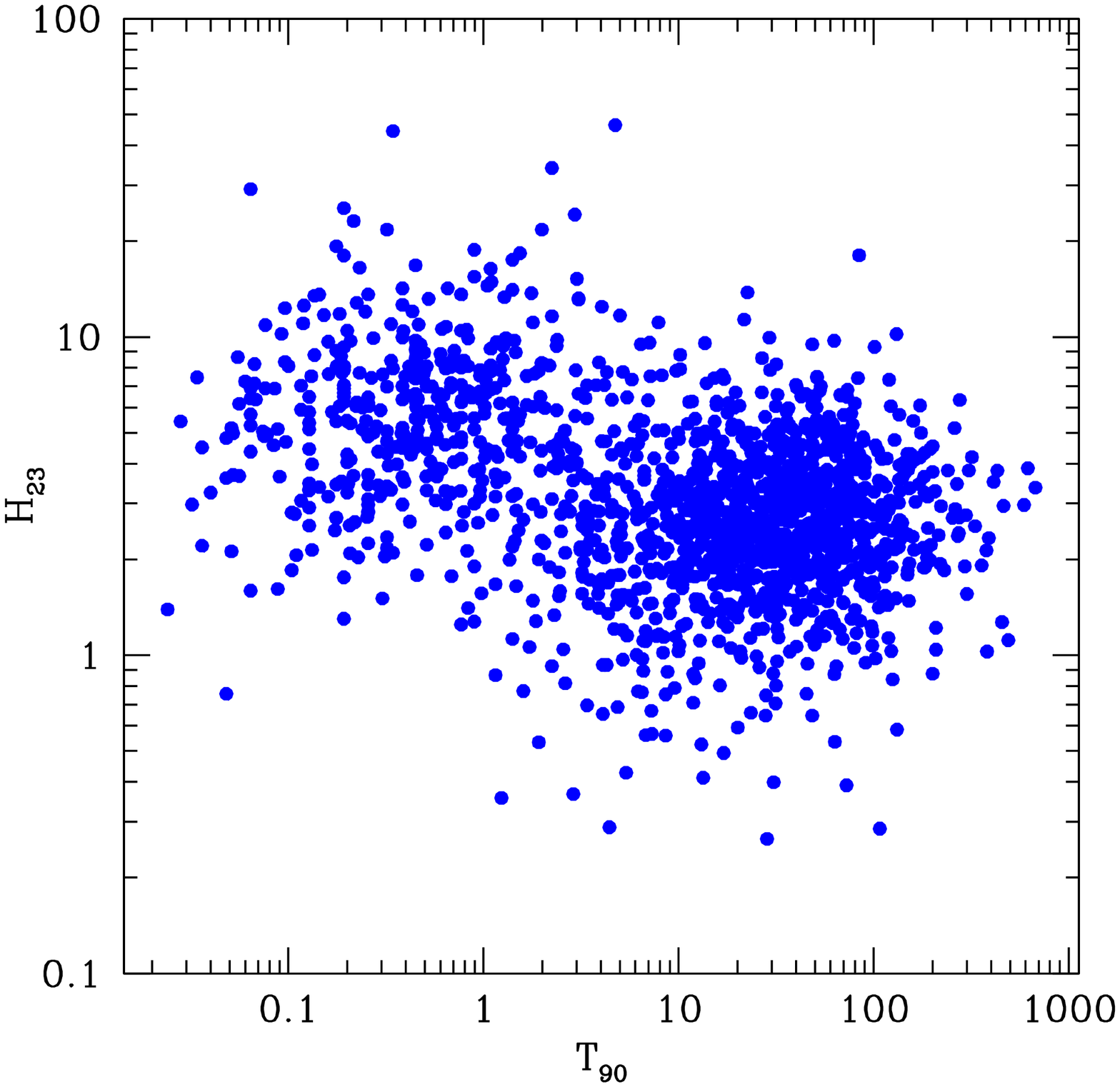}{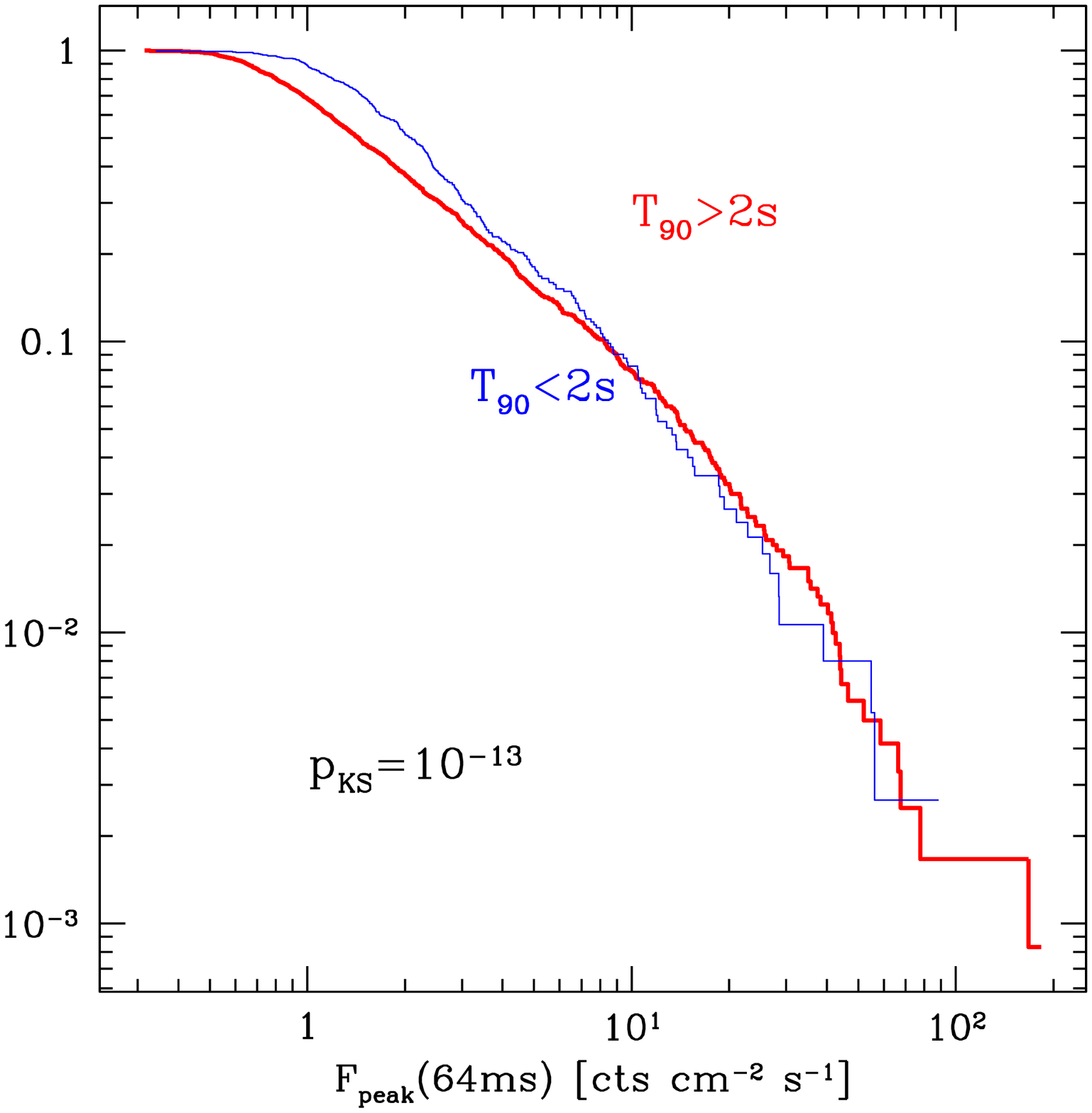}
\caption{The left panel shows 
BATSE gamma ray bursts in the duration - hardness
plane. Two groups of bursts are clearly visible: the long 
and the short burst. The long bursts are typically softer
than the short ones. The right panel shows the peak flux
distributions of the long  ($T_{90}>2s$, the thick line)
and the short bursts ($T_{90}<2s$, the thin line). The two
distributions are different, and assuming that the sensitivity
to both classes of bursts is similar, the probability that they
come from the same distribution is $\approx 10^{-13}$.} 
\label{durhard}
\end{figure}

All that has been written above was based on the assumption that
there is only one population of GRBs. There are however some
indications that there may exist different classes of GRBs.
One of the first BATSE results was to show the bimodality
of the duration distribution \citep{1993ApJ...413L.101K}, with
typical durations of less than a second and a few tens of seconds.
At the same time it has been found that GRBs can be 
grouped according to their duration and variability 
properties and form two classes - short variable, and 
long smooth \citep{1993ApJ...413L..11L}.  Moreover 
brightness distributions of bursts with different hardness ratios
are different \cite{PizzichiniA}. The low hardness ratio bursts 
show a typically steeper behavior in the cumulative brightness
distributions, while the harder bursts are those that actually
show the rollover in this distribution.

In the left panel Figure~\ref{durhard} I present the BATSE bursts in the
duration-hardness plane (hardness is defined as the ratio of
the  fluxes in BATSE channels 2 and 3). The duration bimodality
is clearly seen, moreover the long bursts have typically  softer
spectra than the short ones, although there is a significant
overlap. The properties of GRBs have been
analyzed by dividing them into different classes on  the
hardness duration plot. \citet{1995Ap&SS.231...43B} used a line
defined by $HR=0.5 T_{90}^{1/2}$. \citet{1998ApJ...497L..21T}
divides the bursts into four groups by two lines: duration of 
$2.5\,$s, and hardness ratio $HR=3$.  Only the long hard bursts 
$\log(N)$-$\log(S)$ distribution shows deviation from the
Euclidean $-3/2$ slope. I present the  peak flux distributions
of the long and short burst in the right panel Figure~\ref{durhard}. Such
different brightness distributions could indicate that there are
two populations with different luminosities. On the other hand 
a similar effect has been found when analyzing  distribution the
peaks in BATSE bursts \citep{1997ApJ...489..175P}. 
Peaks in GRBs evolve typically from hard to soft spectra, and 
one could speculate that the short bursts are just single peaked,
while the longer ones contain more of the soft emission from the
tails of the peaks. 
It should be noted that 
SAX probes only long bursts (duration larger than $6$\,s) and so
far all but one burst have been in the hard category.

The discovery of an unusual supernova in the errorbox
of GRB980425 \citep{1998Natur.395..670G} has sparked a debate 
on a possibility of the association of GRBs and supernovae.
If the association  were correct then gamma-ray luminosity of
GRB980425 would be much smaller than the typical luminosity
of GRBs. It is still unclear whether this association is correct
and if there exists a class of supernova related GRBs. If it
does then it would be interesting to find out what fraction of
GRBs this class constitutes, and whether the supernovae related
bursts are related to one of the GRB classes shown above.
If there exists a class of under luminous, supernova related GRBs
then the faint bursts should be dominated by distant bursts in
this class. A future very sensitive experiment may resolve this,
however it looks more promising to pursue better quick
localizations  and multi wavelength followups.

\section{Summary}

Gamma-ray bursts sky distribution is consistent with isotropy to
a very high degree. The estimates of the distance scale and
spatial distribution of GRBs seem to converge. The typical
gamma-ray burst lie at redshifts of $1$ to $2$, however this
distribution probably has a long tail extending to higher
redshifts. GRB luminosity function is broad; from the
observation of just a few bursts we already see that the
observed luminosity function is almost a factor of one thousand
wide. This means that the luminosity function plays a crucial
role in shaping   the GRB brightness distribution. One of the
the brightest bursts seen by BATSE - GRB990123 - lies at the
redshift of $1.60$! It is possible that there exist different
classes of GRB within what we call now cosmic GRBs. This could
reflect either different physical mechanism  connected with the
central engine (compact object mergers or collapsars), or
perhaps even orientation effects. Only the long and hard bursts
duration distribution exhibits a rollover characteristic for the
cosmological models, and this could be an indication that the
remaining bursts with Euclidean brightness distributions are
different. Finally, one should note that while the evidence for
identification of GRB980425 with SN1998bw is still uncertain, it
is certainly worth to investigate a possible GRB-supernova
connection.

\acknowledgments

The author wishes to thank the organizers for making  the
Graftavallen meeting  a success in both science and fun
categories. I would like to thank S.~Bajtlik, K. Belczy{\'n}ski,
and M. Chodorowski for help  in preparing this manuscript. This
work has been supported by the KBN grant 2P03D00415.

%
%
\newcommand{\iaucirc}{IAU Circ}
\newcommand{\nat}{Nature}


\end{document}